\documentclass[aps,superscriptaddress,twocolumn,twoside,floatfix,pra,nofootinbib,a4paper]{revtex4-2}

\usepackage{physics}
\usepackage[charter,cal=cmcal,sfscaled=false]{mathdesign}

\usepackage{dcolumn}% Align table columns on decimal point
\usepackage{bm}% bold math
\usepackage{dsfont}
\usepackage{graphicx,epsfig}
\usepackage{amsmath}
\usepackage{braket}
\usepackage{caption}
\usepackage{float}
\usepackage[shortlabels]{enumitem}

\usepackage{blindtext}
\usepackage{color}

\usepackage[colorlinks]{hyperref}
\hypersetup{
	colorlinks = true,
	urlcolor = {blue},
	citecolor = {blue},
	linkcolor= {blue}
}

\usepackage[bottom]{footmisc}

\renewcommand{\eqref}[1]{Eq.~(\ref{#1})}
\newcommand{\figref}[1]{Fig.~\ref{#1}}

\newcommand{\secref}[1]{Sec.~\ref{#1}}

\begin{document}

\title{Global Restrictions Under Local State Discrimination}

\author{Carles Roch i Carceller}\email{carles.roch\_i\_carceller@teorfys.lu.se}\affiliation{Physics Department and NanoLund, Lund University, Box 118, 22100 Lund, Sweden}
\author{Alexander Bernal}\email{alexander.bernal@csic.es}\affiliation{Instituto de Física Teórica, IFT-UAM/CSIC, Universidad Autónoma de Madrid, Cantoblanco, 28049 Madrid, Spain}
%\author{Hanwool Lee}\affiliation{University of Jyväskylä}

\begin{abstract}
We investigate how local distinguishability can restrict global properties of states shared between two parties. We begin exploring how non-locality becomes limited by local state discrimination and observe a non-trivial trade-off between the Clauser-Horne-Shimony-Holt (CHSH) violation and success probability of local discrimination. We extend our findings to bounding the maximally entangled state fidelity and global observables such as the energy. Our results show that local state discrimination can become a powerful tool to limit global behaviours, e.g.~from entangled adversaries in quantum cryptography.
\end{abstract}

\maketitle

\section{Introduction}

Quantum states, unlike their classical counterparts, are not always perfectly distinguishable by nature. Quantum mechanics fundamentally prohibits the existence of a measurement capable of discriminating any arbitrary ensemble of state preparations, primarily due to the uncertainty principle and the fact that quantum states can have overlapping supports within a shared Hilbert space. This raises the central question in quantum state discrimination: to what extent can quantum measurements distinguish between a given set of state preparations? \cite{Barnett2009,Bae2015}. The inherent indistinguishability of quantum states, embedded within the foundational postulates of quantum mechanics, was a subject of interest well before the advent of quantum information science \cite{helstrom1967,helstrom1968,helstrom1969}. From the fundamental roots of quantum state discrimination in prepare-and-measure scenarios, numerous applications have emerged, spanning communication \cite{Bae2011,Tavakoli2021,Navascues2023}, quantum cryptography \cite{bennett1984,Gisin2002,marcin2011}, randomness certification \cite{li2011,brask2017,Carceller2022} as well as witnessing contextuality \cite{schmidt2018,Flatt2022,Carceller2024} and dimension \cite{Brunner2013}.

On the other hand, quantum states shared between multiple parties can exhibit quantum correlations non-achievable by classical means. Sixty years ago, John Bell demonstrated that the predictions of quantum theory are incompatible with the notion of locality \cite{Bell1964}. Non-local behaviours can be detected through the violation of certain inequalities built from linear functionals of observed probabilities, commonly known as Bell inequalities. Witnessing non-local correlations has enabled important applications in quantum information, most notably allowing to infer inherent properties of physical systems in a device-independent (DI) manner. Noteworthy examples are quantum cryptography \cite{mayers1998,Acin2007}, randomness certification \cite{colbeck2011,acin2012} and self-testing \cite{tsitelson1987}. Non-local behaviours belong to a set of quantum correlations stronger than any classical counterparts, arising from global intrinsic properties of the system, such as entanglement. However, in practical scenarios, directly accessing or measuring these global properties in large quantum systems is challenging and often infeasible. Typically, only local measurements on subsystems or individual particles are experimentally accessible. For instance, in Bell scenarios, a global property of the system (entanglement) is inferred through local operations.

In this work we aim to find a connection between behaviours that might arise from global intrinsic properties of the system, such as entanglement, and local distinguishability of state preparations. In one extreme, if the average of a set of quantum states in a fixed dimension is maximally entangled, all states in the ensemble must be the same maximally entangled state, making all preparations indistinguishable. Oppositely, if the ensemble is separable, there is no restriction in the closeness of all preparations. Intermediate points between maximally entangled and separable ensembles intuitively suggest a trade-off between the level of entanglement 
and local distinguishability \cite{saha2024}. Here we explore how local state discrimination generally restricts global properties in bi-partite systems. We begin looking the simplest case of two bi-partite state preparations, and study their local distinguishability given a certain Bell violation. Next, we extend the scenario to $N$ state preparations globally restricted in a twofold manner: by a bounded fidelity with any maximally entangled state in $N^2$ dimensions; or with a bounded global observable such as the energy. Our findings suggest that global properties of the system can be bounded through local state discrimination. Finally, we end this work comparing the performance of global v.s.~local measurement strategies in terms of state discrimination, finding a gap in their performance when the global ensemble of states becomes allowed for non-local behaviours.  \\

\section{Scenario}

\begin{figure}
    \centering
    \includegraphics[width=0.45\textwidth]{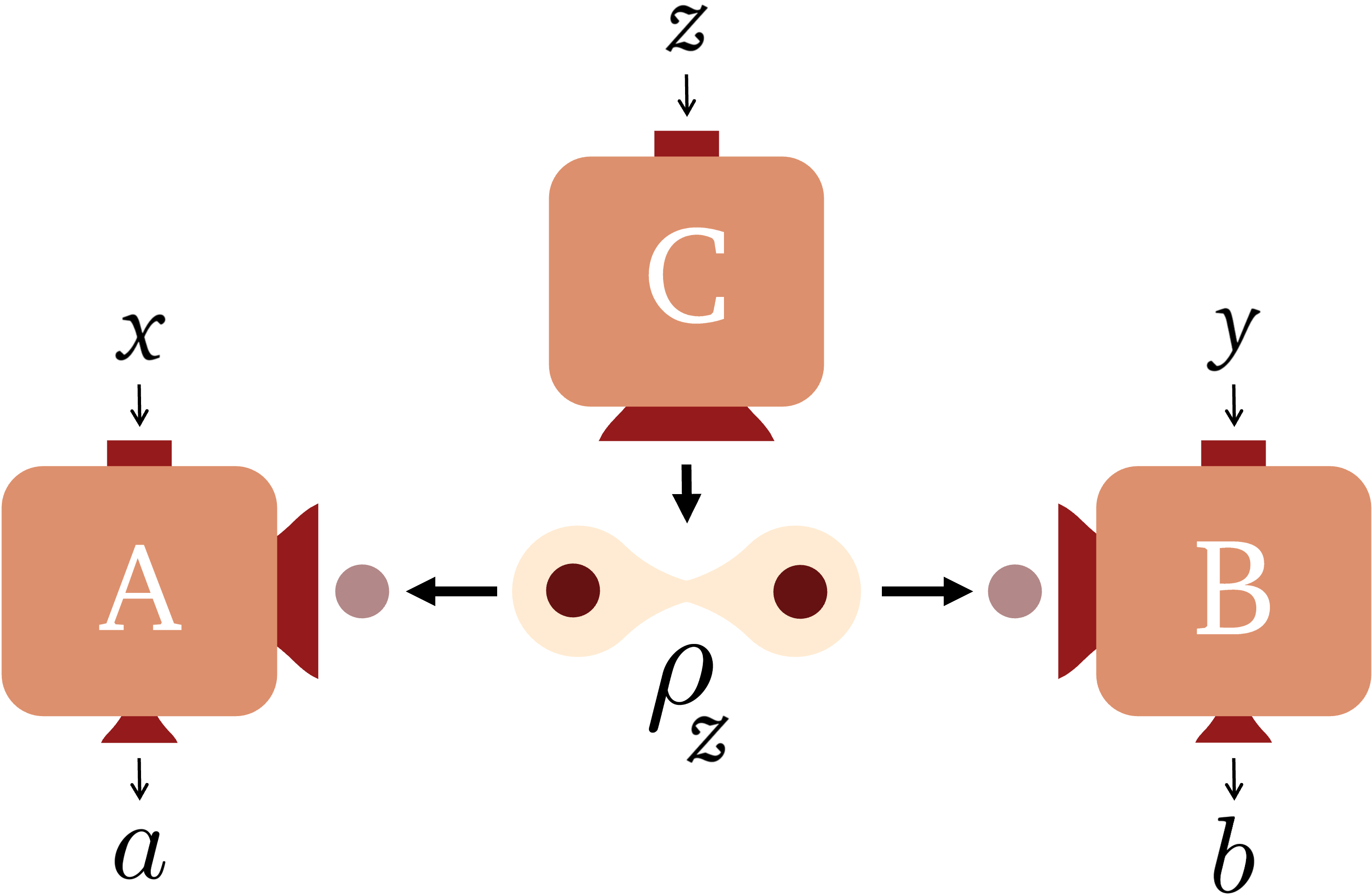}
    \caption{Tri-partite prepare-and-measure scenario. Charlie receives an input $z$ and prepares a quantum state $\rho_z$ which is distributed among Alice and Bob. These receive inputs $x$ and $y$ and spit $a$ and $b$ as measurement outcomes, respectively.}
    \label{fig:scenario}
\end{figure}

Consider the tri-partite scenario in \figref{fig:scenario}. One party, namely Charlie, owns a device that, given a classical input $z=0,\ldots,N-1$, prepares a quantum state $\rho_z$ with prior probability $p_z$. This state is distributed among two distinct and separated systems, namely Alice and Bob, with classical inputs $x$ and $y$ and measurement outcomes $a$ and $b$, respectively. We assume that Alice and Bob have a fixed local dimension $d$. Our first goal is to characterize the local distinguishability of the ensemble $\rho:=\sum_z p_z \rho_z$ in Alice and Bob's side. We denote the local state discrimination success probability as
\begin{align}
\label{eq:ps}
    p_{s}^{\text{L}} := \sum_z p_z \Tr\left[\rho_z \left(A_{z|x^\ast}\otimes B_{z|y^\ast}\right)\right] \ ,
\end{align}
where $\{A_{a|x}\}$ and $\{B_{b|y}\}$ are Alice and Bob's measurement operators, and we fix the settings $x=x^\ast$ and $y=y^\ast$ to distinguish between state preparations. The local success probability in \eqref{eq:ps} collects all events in which both Alice and Bob successfully discriminate the quantum states prepared by Charlie. Alternatively, one might be interested in the individual local success probabilities,
\begin{align}
	p_s^{\rm A} =& \sum_z p_z \Tr\left[\rho_z^A A_{z|x^\ast}\right] \label{eq:psA} \\
	p_s^{\rm B} =& \sum_z p_z \Tr\left[\rho_z^B B_{z|y^\ast}\right] \label{eq:psB} \ ,
\end{align}
for $\rho_z^A = \Tr_B\left[\rho_z\right]$ and $\rho_z^B = \Tr_A\left[\rho_z\right]$ being the traced out portions of the prepared states. Considering individual success probabilities in state discrimination implies a more general perspective. Indeed, all events in which both Alice and Bob successfully discriminate Charlie's states--i.e.~those counted in $p_s^{\rm L}$ from \eqref{eq:ps}--are contained in all events in which either Alice or Bob succeed in the discrimination task, captured in $p_s^{\rm A}$ and $p_s^{\rm B}$ from \eqref{eq:psA} and \eqref{eq:psB} respectively. Mathematically, this relationship is expressed as $p_s^{\rm A}, \ p_s^{\rm B} \geq p_s^{\rm L}$. It turns out, however, that both approaches--i.e.~using $p_s^{\rm L}$ or $p_s^{\rm A}$ and $p_s^{\rm B}$--are equivalent within the scope of our work. Namely, bounding global properties of the averaged ensemble $\rho$ given local observations in Alice and/or Bob's laboratories in the form of state discrimination leads to the saturation $p_s^{\rm A}= p_s^{\rm B} = p_s^{\rm L}$. The characterisation of local state discrimination through $p_s^{\rm L}$ hence becomes a convenient choice. The key idea in the following sections is that bounding $p_s^{\rm L}$ directly constrains how close Charlie's preparations $\rho_z$ are to a specific target state. Moreover, if maximising $p_s^{\rm L}$ one finds the highest achievable $p_s^{\rm A}$ or $p_s^{\rm B}$, one can establish a one-to-one relation between a particular global property and local state discrimination bounds.

In the following, we will stay with $p_s^{\rm L}$ from \eqref{eq:ps} to quantify the distinguishability of state preparations. We will systematically study how a bound on $p_s^{\rm L}$ limits three global properties of the ensemble: non-local correlations, the fidelity with any maximally entangled state and the global energy.

\section{Non-local correlations} \label{sec:III}

We begin studying how non-local correlations can become limited through local state discrimination. First, we need to include a measure of non-locality into the picture. To do so, we will assume that Alice and Bob are able to dedicate the rest of the settings $x$ and $y$ to play a non-local game with the ensemble $\rho$. We take the simplest case with two inputs and two outputs, i.e.~the Clauser-Horne-Shimony-Holt (CHSH) game \cite{CHSH1969}. Assuming all inputs $x,y$ and $z$ are equiprobable, the winning probability of the CHSH game can be written as,
\begin{align}
    p_{\text{win}}^\text{CHSH} := \dfrac{1}{4}\sum_{a,b,x,y}\Tr\left[\rho \left(A_{a|x}\otimes B_{b|y}\right)\right]\delta_{a\oplus b, x\wedge y} \ ,
\end{align}
for $a,b,x,y=\{0,1\}$ and $\delta_{i,j}$ being the Kronecker delta. If $p_{\text{win}}^\text{CHSH}\leq 0.75$, the ensemble $\rho$ can be separable and, with no further restrictions, fully distinguishable in Alice and Bob's sides. Otherwise, this is no longer true. In fact, Tsirelson showed in Ref.~\cite{Tsirelson1980} that the maximal $p_{\text{win}}^\text{CHSH}$ in a quantum model is given by $\frac{1}{2}(1+\frac{1}{\sqrt{2}})$ for Bell state preparations $\{\ket{\phi^{\pm}},\ket{\psi^{\pm}}\}$. Therefore, if Alice and Bob saturate the quantum bound, and their local dimension is limited to a qubit space, the ensemble $\rho$ must be one of the Bell states. This makes Charlie's states completely indistinguishable.

Let us investigate further the relation between the distinguishability of the prepared ensemble and its level of entanglement given by a particular violation of the CHSH game. Since we are interested in studying ensembles $\rho$ able to violate the CHSH inequality, we will consider both state preparations to be pure, $\rho_z=\ketbra{\psi_z}$ and symmetrically close to one of the Bell states, say $\ket{\phi^+}$. Additionally, we can w.l.g.~assume that both pure state preparations span an effective qubit-space given by any orthonormal basis, say $\{\ket{\phi^+},\ket{\phi^-}\}$. This allows us to write the state preparations as
\begin{align}
\label{eq:states}
    \ket{\psi_z} = \sqrt{\frac{1+\delta}{2}}\ket{\phi^+} + (-1)^z \sqrt{\frac{1-\delta}{2}}\ket{\phi^-} \ .
\end{align}
With this construction, the overlap of both states is $\braket{\psi_0|\psi_1}=\delta$, and their fidelity with the Bell state $\ket{\phi^+}$ is $|\braket{\psi_z|\phi^+}|^2=(1+\delta)/2$, $\forall z$. 

Let us look now at the maximal $p_{\text{win}}^\text{CHSH}$ reachable with the given ensemble consisting of equiprobable preparations. Assume that Alice performs the measurements $A_{0|0}=\ketbra{0}$, $A_{1|0}=\ketbra{1}$ for $x=0$ and $A_{0|1}=\ketbra{+}$, $A_{1|1}=\ketbra{-}$ for $x=1$. In Bob's side we will allow him to perform the measurements parametrized with the angle $\theta$,
\begin{align}
    \Pi\left(\theta\right):= \frac{1}{2}\left[\mathds{1} + \sin\theta\sigma_x + \cos\theta\sigma_z \right] \ ,
\end{align}
such that $B_{0|0}=\Pi(\theta)$, $B_{1|0}=\Pi(\theta+\pi)$ for $y=0$ and $B_{0|1}=\Pi(-\theta)$, $B_{1|1}=\Pi(-\theta+\pi)$ for $y=1$. Using these measurements with the state preparations in \eqref{eq:states}, the probability of winning the CHSH game becomes $p_{\text{win}}^\text{CHSH}=(2+\cos\theta+\delta\sin\theta)/4$. One can analytically find that this probability reaches its maximum whenever $\tan\theta=\delta$, which reads
\begin{align}
    p_{\text{win}}^\text{CHSH} \leq \frac{1}{4}\left(2+\sqrt{1+\delta^2}\right) \ . \label{eq:chshbound}
\end{align}
As we detail in \cite{supp}, this bound, although derived for the states in \eqref{eq:states}, is general for any set $\{\ket{\psi_z}\}_{z=0,1}$ of equiprobable states and can be further generalized for the non-equiprobable scenario.
Hence, whenever \eqref{eq:chshbound} is saturated, the fidelity of $\rho$ with any maximally entangled state cannot be lower than $(1+\delta)/2$, which means that the overlap of the pure states $\rho_z$ that form the ensemble cannot be smaller than $\delta$.

Next, we study the distinguishability of $\rho_z$ through the maximal success probability for Alice and Bob using local measurement strategies. Their respective local states, $\rho_z^A=\Tr_{B}\left[\rho_z\right]$ and $\rho_z^B=\Tr_{A}\left[\rho_z\right]$, are in both sides equivalent to
\begin{align} \label{eq:local_qubits}
    \rho_z^{A} = \rho_z^{B} = \frac{1}{2}\left[\mathds{1}+(-1)^z \sqrt{1-\delta^2}\sigma_z\right] \ .
\end{align}
These states can be optimally discriminated using local projective measurements. Namely, Alice and Bob perform the projective measurements $A_{0|x^\ast}=B_{0|y^\ast}=\ketbra{0}$ and $A_{1|x^\ast}=B_{1|y^\ast}=\ketbra{1}$. These allow them to reach the maximum success probability,
\begin{align}\label{eq:psLCHSH}
    p_{s}^{\text{L}}\leq\frac{1}{2}(1+\sqrt{1-\delta^2})\ ,
\end{align}
for two pure qubit states with overlap $\delta$, also known as the Helstrom bound \cite{helstrom1969}. This bound is the maximum success probability attainable by global measurements. Since any success probability attained by local measurements must be lower or equal than the maximum global success probability, this represents the best measurement strategy that can be performed locally. Furthermore, the inequality in \eqref{eq:psLCHSH} is saturated when $p_{s}^{\text{A}}=p_{s}^{\text{B}}=p_{s}^{\text{L}}$.

The interesting part arises when we consider the statement in reverse. Specifically, suppose Alice receives two quantum states $\rho_z^{A}$ for $z=0,1$. Alice can measure these states using a device programmed to optimally distinguish them. Now, Alice wants to know if her set of states might be correlated with states owned by another party, Bob. If Alice's states $\rho_z^{A}$ are pure, any global states $\rho_z$ shared between her and Bob will be separable. In the worst case--i.e. when $\rho$ is maximally entangled--her states $\rho_z^{A}$ are completely mixed, having only non-zero weights on the diagonal elements. All Alice knows for sure is that she can optimally distinguish between the states $\rho_z^{A}$	with a success probability $p_s^{\text{L}}$. This knowledge directly implies an upper bound on the overlap $\braket{\psi_0|\psi_1}\leq\delta$ via the Helstrom bound in \eqref{eq:psLCHSH}. Namely,  $\delta^2 = 1- (2p_s^{\text{L}}-1)^2$ or equivalently that Alice's states $\rho_z^{A}$ are at worst similar to those in \eqref{eq:local_qubits}.  Moreover, this means Alice can be confident that her states may be part of a shared bi-partite ensemble with Bob, where the global correlations are constrained to win the CHSH game with probability bounded by \eqref{eq:chshbound}, or in terms of the local success probability, 
\begin{align}\label{eq:chshboundlocal}
    p_{\text{win}}^\text{CHSH} \leq \frac{1}{4}\left(2+\sqrt{2-(2p_s^{\text{L}}-1)^2}\right) \ .
\end{align}
Through local state discrimination, Alice can therefore infer limits on the potential global correlations.

\begin{figure}
    \centering
    \includegraphics[width=0.5\textwidth]{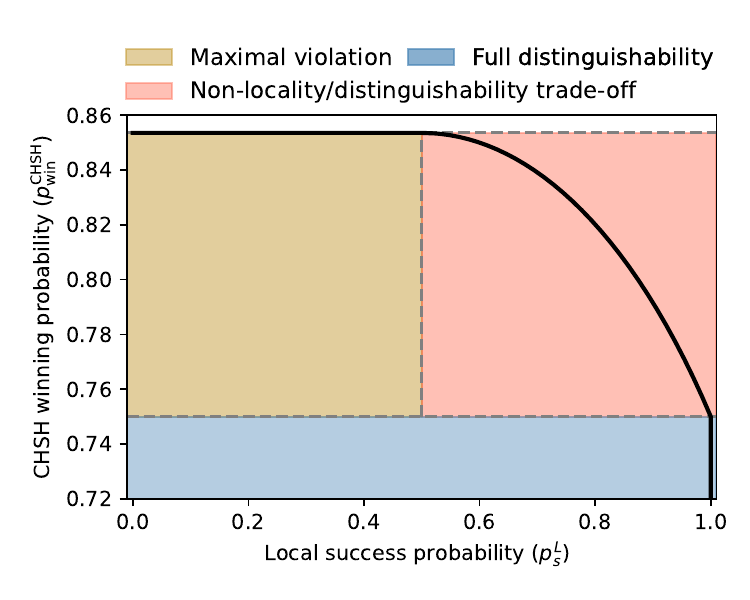}
    \caption{Trade-off between CHSH violation and local state distinguishability characterized through the state discrimination success probability.}
    \label{fig:tradeoff}
\end{figure}

We additionally constructed a numerical tool based on semidefinite programming (SDP) \cite{tavakoli2024}. This method is used to numerically compute a bound on $p_{\text{win}}^\text{CHSH}$ for any set of local measurements in Alice and Bob sides with a given ensemble with an overlap bounded by $\delta$. The SDP operates in a \textit{see-saw} manner. Namely, it fixes random measurements on Alice side and optimizes $p_{\text{win}}^\text{CHSH}$ for any set of measurements $\{B_{b|y}\}$ in Bob's side, i.e.~POVMs restricted by positivity $B_{b|y}\succeq 0$ and completeness $\sum_b B_{b|y}= \mathds{1}$. Once a solution is found, Bob's measurements are then fixed and the optimization runs again through Alice's measurements $\{A_{a|x}\}$ constrained by  $A_{a|x}\succeq 0$ and $\sum_a A_{a|x}= \mathds{1}$. The optimization iterates upon a certain level of convergence. Our numerical findings agree with our analytical bound from \eqref{eq:chshboundlocal} up to numerical precision and are plotted in \figref{fig:tradeoff}. 

For interested readers, we refer them to \cite{supp} where we derive bounds on the CHSH winning probability analogous to \eqref{eq:chshbound} but with $N=3,4$ preparations and local qubit dimension.
\\

\section{Maximally entangled state fidelity}

Our findings so far suggest that global properties of an ensemble of bi-partite states $\rho$ can be bounded by local state discrimination. E.g., with $N=2$ preparations, non-local behaviours are limited by local dimensionality and distinguishability. Here, we generalize our findings to a scenario with $N$ bi-partite pure state preparations. In this case, we are no-longer interested in determining the level of non-locality though a Bell inequality, but more specifically we aim to bound the fidelity with any maximally entangled state in dimension $N^2$. In order to illustrate this, we construct an explicit ensemble of states equally distant to a maximally entangled state in an $N^2$-dimensional space. Later on we will justify the generality of the results derived by using this family. For instance, let us take the set of generalized maximally entangled states \cite{bennet1993}
\begin{align}
    \ket{\xi_{n}} = \frac{1}{\sqrt{N}}\sum_{j}e^{i\frac{2\pi jn}{N}}\ket{j}\otimes\ket{j}\label{eq:generalised_basis} \ ,
\end{align}
for $n=0,\ldots,N-1$ as an effective basis of the $N$-dimensional space spanned by the $N$ pure preparations. In this basis, we construct the set of $N$ pure states to be equidistant to one maximally entangled state of the chosen basis, say $\ket{\xi_0}$. Namely,
\begin{align}
    \!\!\!\ket{\psi_z}\!=\! \sqrt{\frac{1\!+\!(N\!-\!1)\delta}{N}}\ket{\xi_0} + \sqrt{\frac{1-\delta}{N}}\sum_{j=1}^{N-1}e^{i\frac{2\pi jz}{N}} \ket{\xi_j} . \label{eq:qutrit_states} 
\end{align}
The ensemble built from their statistical mixture, $\rho=\sum_z p_z \ketbra{\psi_z}$ with $p_z=1/N$, belongs to a well-studied family of states known as axisymmetric states \cite{Eltschka2013,Eltschka2015,Seelbach2022}. In the following we use a set of properties of states in \eqref{eq:qutrit_states}, explained in more detail in \cite{supp}. For instance, all states are equally spaced with the same pair-wise overlap $\braket{\psi_z|\psi_{z'}}=\delta$, $\forall z,z'$. Furthermore, their partial-trace is diagonal and equivalent in Alice and Bob's side, i.e.~$\rho_z^A=\rho_z^B=\text{diag}(\lambda_z^0,\ldots,\lambda_z^{N-z},\ldots,\lambda_z^{N-1})$, for $\lambda_z^{N-z}=p_{s}^{N}(\delta)$ and $\lambda_z^{i}=p_{e}^{N}(\delta)$, $\forall i\neq N-z$, with $N-0\equiv0$. We have denoted by $p_{s}^{N}(\delta)$ the maximal state discrimination success probability with $N$ equidistant pure states \cite{pauwels2024},
\begin{align}
p_{s}^{N}(\delta)\!=\!\frac{1}{N^2
}\left(\sqrt{1+(N-1)\delta}+(N-1)\sqrt{1-\delta}\right)^2 ,
\end{align}
and by $p_{e}^{N}(\delta)=\frac{1-p_s^N(\delta)}{N}$ the minimum error probability. Therefore, by using the ensemble in \eqref{eq:qutrit_states}, Alice and Bob can both optimally discriminate their respective shares by projecting onto the largest eigenvalue of the states they receive. 

This observation leads to the following statement: if a success probability $p_s^{\text{L}}$ is observed,  we can always associate an overlap $\delta$ to the ensemble via $p_s^{\text{L}}=p_{s}^{N}(\delta)$.  In particular,  due to the optimality of $p_{s}^{N}(\delta)$ the overlaps between all global state preparations $\braket{\psi_z|\psi_{z'}}$ cannot be greater than $\delta$. 

To see why, take an ensemble of $N$ equidistant pure states $\{\ket{\psi_z}\}_z$ with overlaps $\braket{\psi_z|\psi_{z'}}=\delta$, $\forall z,z'$. Any other ensemble of $N$ pure states $\{\ket{\phi_z}\}_z$ with $\braket{\phi_z|\phi_{z'}}\geq\delta$, $\forall z,z'$, is necessarily less or equally distinguishable. That is,
\begin{align}
\braket{\phi_z|\phi_{z'}}\geq\delta \ \forall z,z' \ \Rightarrow \ p_{s}^{N}(\delta) \geq p_s^{\phi} \ ,
\end{align}
where $p_s^{\phi}$ is the maximum success probability in discriminating the states from the ensemble $\{\ket{\phi_z}\}_z$. Hence, to find a bound on the success probability, it is enough to consider the limiting case of ensembles with equidistant states.

Now suppose we observe $p_{s}^{N}(\delta)$ for the ensemble $\{\ket{\psi_z}\}_z$ with overlaps $\braket{\psi_z|\psi_{z'}}=\delta$, $\forall z,z'$. This success probability is also attainable for ensembles of pure states $\{\ket{\varphi_z}\}_z$ with $\braket{\varphi_z|\varphi_{z'}}\leq\delta$, $\forall z,z'$. Specifically, if we modify the ensemble $\{\ket{\psi_z}\}_z$ by reducing some of the overlaps--such as by separating one state from the rest arriving to the $\{\ket{\varphi_z}\}_z$ configuration--the overall distinguishability increases. In this case,  $p_{s}^{N}(\delta)$ remains achievable with a non-optimal measurement. This implies
\begin{align}
p_s^{\varphi} = p_{s}^{N}(\delta) \ \Rightarrow \ \braket{\varphi_z|\varphi_{z'}}\leq\delta \  \ \forall z,z' \ ,
\end{align}
which is our original statement. Furthermore, our global construction in \eqref{eq:qutrit_states} corresponds to a set of equidistant pure states with $\braket{\psi_z|\psi_{z'}}=\delta$, and their traced-out portions are already saturating optimal local state discrimination. Then again, this constitutes the best measurement strategy that can be implemented locally, extending our results from \secref{sec:III} to arbitrary $N$ pure states.

Our formulation hinges that local state discrimination is related to the closeness of global pure states parametrized through their overlaps. These can be directly linked to the maximal fidelity of the ensemble $\rho$ with any maximally entangled state. Let us define
\begin{align}
f_{\text{max}}(\rho) = \underset{\Psi}{\max} \bra{\Psi}\rho\ket{\Psi}
\end{align}
for $\Psi$ being any maximally entangled bi-partite state with local dimension $N$. 
Now, consider that Alice is given the task of determining $f_{\text{max}}(\rho)$ only knowing how well can she discriminate her local shares of the states $\rho_z$ forming the ensemble $\rho$. 
Knowledge on $p_s^{\text{L}}$ implies a closeness limitation between all global states forming the ensemble, in the form of limited overlaps. Moreover, in \cite{supp} we extend this to a bound on the fidelity of the ensemble with any other global pure state, e.g.~any maximally entangled state, leading to
\begin{align}
\!\!\!f_{\text{max}}(\rho)\!\leq\!\frac{1\!+\!(N\!-\!2)(1\!-\!p_s^{\text{L}})\!+\!2 \sqrt{(N\!-\!1)(1\!-\!p_s^{\text{L}})p_s^{\text{L}}}}{N}. \label{eq:max_fid_bound}
\end{align} 
Here again, we see that the maximum entangled state fidelity becomes bounded by how distinguishable are the state preparations locally.
\\

\section{Local distinguishability limited by global energy}

Once we determined that local statistics constrain global state properties, we continue identifying global observables that find similar limitations. Let $H$ be a global observable such that $\Tr\left[\rho H\right] \geq \alpha$. We choose this observable to be the global energy of the system, i.e. $H=\mathds{1} - \ket{\text{vac}}\bra{\text{vac}}$, for $\ket{\text{vac}}$ being the vacuum component. Conceptually, measuring $H$ corresponds, e.g., to simply using a power-meter to assess the average power in optical states. 

Let us now relate the bound $\alpha$ with our findings. In \eqref{eq:max_fid_bound} we showed that a bounded local success probability $p_s^{\text{L}}$ guarantees a bound on the closeness of the global ensemble $\rho=1/N\sum_z\ketbra{\psi_z}$ to a maximally entangled state in terms of a bound on their overlap $\delta$. This closeness relation works with any pure global state (see \cite{supp}), in particular with the vacuum $\ket{\text{vac}}$. For instance, take the set of states in \eqref{eq:qutrit_states} and unitarily rotate them to be all equally close to $\ket{\text{vac}}$. We know that the fidelity of each state $\ket{\psi_z}$ with the vacuum will be the same as the fidelity of the ensemble $\rho$ with the vacuum, i.e.~$|\braket{\psi_z|\text{vac}}|^2=\bra{\text{vac}}\rho\ket{\text{vac}}\leq\frac{1+(N-1)\delta}{N}$. This implies that the bound on the global pair-wise overlaps $\delta$ can be directly translated to the following global energy bound
\begin{align}
\Tr\left[\rho H\right]\geq\left(1-\frac{1}{N}\right)(1-\delta)=\alpha \ .
\end{align}
We can bring this global property to a local observable through the maximum success probability. Indeed, one finds that
\begin{align}
p_s^{\text{L}}\leq \frac{1}{N^2}\left(\sqrt{N(1-\alpha)}+\sqrt{N(N-1)\alpha}\right)^2 \ .
\end{align}
Any local state discrimination success probability that satisfies this inequality certifies that the global energy cannot be lower than $\alpha$. 

Similar bounds can be analogously found for arbitrary global observables that can be expressed in terms of rank-$1$ operators. \\

\begin{figure*}[t]
    \centering
    \includegraphics[width=0.9\textwidth]{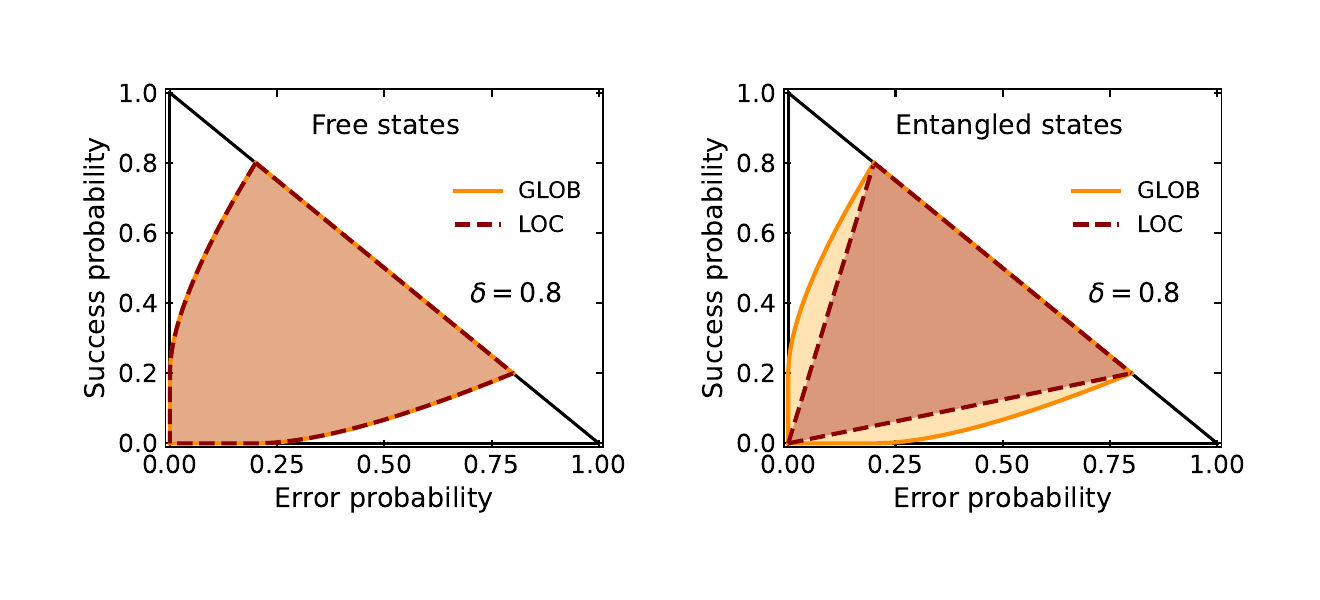}
    \vspace{-1cm}
    \caption{Observable success and error probabilities in a bi-partite two-state discrimination scenario. Global (GLOB) measurement strategies are able to perform better than local (LOC) strategies whenever state preparations are entangled in the form of \eqref{eq:states}.}
    \label{fig:ps_pe}
\end{figure*}

\section{Global vs.~local measurements}

We now turn towards comparing the performance of global and local measurement strategies on the task of successfully discriminating the states $\rho_z$. For generality, we consider imperfect experimental devices which might yield inconclusive measurement outcomes labelled by $a,b=\varnothing$ with probability 
\begin{align}
    \label{eq:inc}
    p_{\varnothing}^{\text{L}} := \sum_z p_z \Tr\left[\rho_z \left(A_{\varnothing|x^\ast}\otimes B_{\varnothing|y^\ast}\right)\right] \ .
\end{align}
To study the distinguishing capabilities of local measurement strategies, our goal will be to maximize the success probability $p_{s}^{\text{L}}$ from \eqref{eq:ps} given a bounded rate of inconclusive events $p_{\varnothing}^{\text{L}}\geq p_{\varnothing}$. This optimization shall run over all possible pure state preparations $\rho_z=\ketbra{\psi_z}$ with overlaps bounded by $\braket{\psi_z|\psi_{z'}}\geq\delta$ and measurements strategies $A_{a|x}$ and $B_{b|y}$ in Alice and Bob's devices respectively. For this purpose, we will employ an SDP hierarchy based on building a Gramm matrix \cite{wang2019}. Namely, we define the list of operators $S=\left\{\ket{\psi_z},\left(A_{a|x}\otimes\mathds{1}\right)\ket{\psi_z},\left(\mathds{1}\otimes B_{b|y} \right)\ket{\psi_z}\right\}$. Then, we build the Gramm matrix $G_{u,v}=u^\dagger v$, for $u,v\in S$. By construction, for any set of valid states $\left\{\ket{\psi_z}\right\}$ and POVMs $\left\{A_{a|x}\right\}$, $\left\{B_{b|y}\right\}$, the Gramm matrix $G$ is positive-semidefinite. Reversely, a properly constructed $G\succeq 0$ (that is in this case, with normalised operators $\left\{A_{a|x}\right\}$, $\left\{B_{b|y}\right\}$) yields correlations fitting an outer approximation of the quantum set \cite{navascues2008}. Indeed, note that all observable probabilities can be drawn from the elements $p(a,b|x,y,z)=\bra{\psi_z}A_{a|x}\otimes B_{b|y}\ket{\psi_z}$. Additionally, distinguishability can be operationally constrained implying a bound on the overlaps $\braket{\psi_z |\psi_{z'}}\geq \delta$. Through this SDP hierarchy, we compute the maximum success probability $p_s^{\text{L}}$ given a set of global state preparations with bounded pair-wise overlaps and limited probability of inconclusive events. 

We compare the computed bounds with the equivalent scenario but allowing Alice and Bob to perform a global measurement. That is, consider now that Charlie's preparations are sent to a single device which performs a global measurement described by the POVM $\{M_c\}$ yielding $c$ as measurement outcomes. In this generic prepare-and-measure scenario, we aim to maximise the success probability which can be written as
\begin{align}
    p_s^{\text{G}} = \sum_z p_z \Tr\left[\rho_z M_z\right] \ .
\end{align}
The probability of having inconclusive results can similarly be expressed as
\begin{align}
    p_{\varnothing}^{\text{G}} = \sum_z p_z \Tr\left[\rho_z M_{\varnothing}\right] \ .
\end{align}
Since we are considering equally distributed pure state preparations, i.e. $\rho_z=\ketbra{\psi_z}$ with $\braket{\psi_z|\psi_{z'}}\geq\delta$, $\forall z\neq z'$, we can w.l.g.~bound the dimension of the whole system to be equivalent to the number of preparations. The optimisation in this case turns therefore into a simple state discrimination problem. Given a set of pure states $\rho_z$ equally spaced, we maximise $p_s^{\text{G}}$, for all possible valid POVMs $\{M_c\}$ such that the rate of inconclusive events is bounded by $p_{\varnothing}^{\text{G}} \geq p_{\varnothing}$.

In \figref{fig:ps_pe} we show the observable success $p_s$ and error $p_e=1-p_s-p_{\varnothing}$ probabilities for local (LOC) and global (GLOB) measurement strategies, and state preparations with an overlap bounded by $\delta=0.8$. We additionally computed the same scenario with different values of $\delta$, and higher number of state preparations $N$. In all cases, our observations indicate that local measurement strategies can perform as well as global strategies. Namely, the set of feasible success and error probabilities is equivalent in both cases. This indicates that any other strategy involving local operations with classical communication (LOCC) or separable (SEP) measurements will yield the exact same optimal distributions. It is well known that strict inclusions of behaviours generally hold \cite{bennett1999}, i.e.
\begin{align}
\text{LOC} \subset \text{LOCC} \subset \text{SEP} \subset \text{GLOB} \ .
\end{align}
In the task of discriminating quantum states with bounded overlaps, the optimal correlations can be already satisfied with local operations. Therefore, there is no advantage in using global strategies for state discrimination with only a bound on the overlaps.

This is no longer true if entanglement is restricted. For instance, let us go back to the case of two entangled state preparations $z=0,1$ from \eqref{eq:states}. In the case of having non-zero inconclusive events $p_{\varnothing}^{\text{L}}\geq p_{\varnothing}$, the local success probability needs to satisfy $p_{s}^{\text{L}}\leq\frac{1-p_{\varnothing}}{2}(1+\sqrt{1-\delta^2})$ \cite{Carceller2024}. In \figref{fig:ps_pe} we show the accessible probability distributions if the ensemble $\rho$ is able to violate the CHSH inequality for two quantum states with non-zero overlap. Now, we see that local and global measurement strategies differ in performance. Whenever state preparations are entangled, global measurements turn advantageous for state discrimination. More interestingly, consider noisy state preparations, namely $\rho_z = \nu \ketbra{\psi_z} + (1-\nu)\mathds{1}/4$. Note that whenever both states are identical with $\delta=1$, the parameter $\nu$ corresponds to the visibility of the Werner state \cite{werner1989}. Following the optimal qubit measurement strategy from Ref.~\cite{Carceller2024} we find the critical visibility at which separable measurements can distinguish entangled and non-entangled states is $\nu_{\text{c}}=\frac{1}{\sqrt{1+\delta^2}}$, which can be reached for inconclusive rates $p_{\varnothing}^{\text{L}}\leq \frac{\delta}{\sqrt{1+\delta^2}}$.  \\

\section{Conclusions}

We studied the relationship between local state discrimination and global properties of a shared ensemble of states. Specifically, we unveiled a non-trivial relationship between the achievable CHSH violation with the maximum success probability of distinguishing the state preparations. Our results suggest that one can infer a limitation on non-local behaviours through only accessing local state discrimination. In addition, we looked at the generalised case of $N$ distinct preparations and found that local state discrimination bounds can be used to infer bounds on the maximal fidelity with any maximally entangle state, and also on the global energy of the ensemble. The findings reported in this work indicate that global correlations can be constrained through local observables, potentially offering valuable applications in the fields of quantum communication and cryptography. Further exploration of bi-partite scenarios involving multiple inputs in state preparation might potentially reveal intricate properties related to exotic non-local behaviours, as well as non-trivial distinctions between classical and quantum advantages within prepare-and-measure frameworks. \\

\begin{acknowledgments}
Alexander Bernal acknowledges the hospitality of the Lund quantum information group. We thank Hanwool Lee and Jef Pauwels for discussions. We are also grateful to Armin Tavakoli and Gabriele Cobucci for their valuable input. Carles Roch i Carceller is supported by the Wenner-Gren Foundations. Alexander Bernal is supported through the FPI grant PRE2020-095867 funded by MCIN/AEI/10.13039/501100011033. \\
\end{acknowledgments} 

\bibliography{main}% Produces the bibliography via BibTeX. 

\begin{widetext}
\newpage
\appendix

\section{CHSH optimal violation}\label{ap:chsh}

In this part of the supplemental material we show that the bound in \eqref{eq:psLCHSH} from the main text, attained for the family of states in \eqref{eq:states}, is actually the largest winning probability for the CHSH game given probabilities and states $\{p_z,\ket{\psi_z}\}_{z=0,1}$ verifying $\braket{\psi_0|\psi_1}=\delta$. 

Firstly, we notice that the eigenvalues of the state $\rho=\sum_{z=0}^1 p_z \ketbra{\psi_z}$ are fixed due to the requirement $\braket{\psi_0|\psi_1}=\delta$. Namely, since it is a rank-2 matrix, only two of the eigenvalues will be different from zero. These eigenvalues are computed by solving the characteristic polynomial of the non-trivial $2\times2$ block,
\begin{equation}
    \lambda^2-\Tr\left[\rho\right]\lambda+\frac{1}{2}\left(\Tr\left[\rho\right]^2-\Tr\left[\rho^2\right]\right)=0\ .
\end{equation}
In the case in hand, $\Tr\left[\rho\right]=1$ and $\Tr\left[\rho^2\right]$ is fixed by $\delta$ and $\{p_z\}$,
\begin{equation}
    \Tr\left[\rho^2\right]=\sum_{z,z'=0}^1 p_z p_{z'}\abs{\braket{\psi_z|\psi_{z'}}}^2=\sum_z p_z^2+\sum_{z\neq z'} p_z p_{z'} \delta^2=1-2\sum_{z<z'} p_z p_{z'} (1-\delta^2)\ .
\end{equation}
Let us call $P=\sum_{z<z'} p_z p_{z'}$, then the characteristic polynomial reads
\begin{equation}
    \lambda^2-\lambda+P(1-\delta^2)=0\ ,
\end{equation}
whose solutions are
\begin{equation}
    \lambda_{\pm}=\frac{1}{2}\left(1\pm\sqrt{1-4P(1-\delta^2)}\right)\ .
\end{equation}
Secondly, as shown in Ref.~\cite{verstraete2002}, the optimal CHSH violation among all states $\rho$ with fixed spectrum is given by the so-called Bell diagonal states, i.e.~states whose eigenvectors are Bell states. Moreover, the optimal violation, in terms of the operator form of the CHSH game, is 
\begin{equation}
    \beta=2\sqrt{2}\sqrt{(\lambda_1-\lambda_4)^2+(\lambda_2-\lambda_3)^2}\ ,
\end{equation}
where the eigenvalues are labelled such that $\lambda_1\geq\lambda_2\geq\lambda_3\geq\lambda_4$. Applying it to our scenario, we obtain that the optimal violation is 
\begin{equation}
    \beta=2\sqrt{2}\sqrt{\lambda_1^2+\lambda_2^2}=2\sqrt{2}\sqrt{\Tr\left[\rho^2\right]}=2\sqrt{2-4P(1-\delta^2)}\ .
\end{equation}
In terms of the winning probability of the CHSH game, $p_{\text{win}}^\text{CHSH}=\frac{1}{4}(2+\frac{\beta}{2})$,
\begin{align}
    p_{\text{win}}^\text{CHSH} \leq \frac{1}{4}\left(2+\sqrt{2-4P(1-\delta^2)}\right) \ ,
\end{align}
coinciding with \eqref{eq:psLCHSH}.

Actually, this construction works in the general case of having $N$ state preparations $\{\ket{\psi_z}\}_{z=0,\dots,N-1}\in\mathcal{C}^2\otimes\mathcal{C}^2$ with $\{p_z\}_{z=0,\dots,N-1}$ prior probabilities, for $N=2,3,4$ and $\braket{\psi_z|\psi_{z'}}=\delta\ \ \forall z\neq z'$. In this setup, one needs to compute the traces $\Tr\left[\rho^2\right],\Tr\left[\rho^3\right],\Tr\left[\rho^4\right]$ and find the solutions of the characteristic polynomial,
\begin{equation}
\begin{aligned}
    &\lambda^4-\Tr\left[\rho\right]\lambda^3+\frac{1}{2}\left(1-\Tr\left[\rho^2\right]\right)\lambda^2-\frac{1}{6}\left(1+2\Tr\left[\rho^3\right]-3\Tr\left[\rho^2\right]\right)\lambda\\
    &+\frac{1}{24}\left(1-6\Tr\left[\rho^2\right]+8\Tr\left[\rho^3\right]+3\Tr\left[\rho^2\right]^2-6\Tr\left[\rho^4\right]\right)=0.
\end{aligned}
\end{equation}
For equiprobable state preparations, $p_z=\frac{1}{N}$, the traces are
\begin{equation}
\begin{aligned}
    \Tr\left[\rho^2\right]=&1-\frac{1}{N^2}(1-\delta^2)N(N-1)\ ,\\
    \Tr\left[\rho^3\right]=&1-\frac{1}{N^3}\left[(1-\delta^3)N(N-1)(N-2)+3(1-\delta^2)N(N-1)\right]\ ,\\ 
    \Tr\left[\rho^4\right]=&1-\frac{1}{N^4}\left\{(1-\delta^4)\left[N(N-1)(N-2)(N-3)+2N(N-1)(N-2)+N(N-1)\right]\right.\\
    &\left.+(1-\delta^3)4N(N-1)(N-2)+(1-\delta^2)\left[4N(N-1)+2N(N-1)\right]\right\}\ ,
    \end{aligned}
\end{equation}
leading to the eigenvalues
\begin{equation}
    \begin{aligned}
        N=2:\quad& \lambda\in\left\{0,0,\frac{1}{2}\left(1-\delta\right),\frac{1}{2}\left(1+\delta\right)\right\}\ ,\\
        N=3:\quad& \lambda\in\left\{0,\frac{1}{3}\left(1-\delta\right),\frac{1}{3}\left(1-\delta\right),\frac{1}{3}\left(1+2\delta\right)\right\}\ ,\\
        N=4:\quad& \lambda\in\left\{\frac{1}{4}\left(1-\delta\right),\frac{1}{4}\left(1-\delta\right),\frac{1}{4}\left(1-\delta\right),\frac{1}{4}\left(1+3\delta\right)\right\}\ .\\
    \end{aligned}
\end{equation}
Hence, the optimal violation reads
\begin{equation}
    \begin{aligned}
        N=2:\quad& \beta=2\sqrt{1+\delta^2}\ ,\\
        N=3:\quad& \beta=\frac{2\sqrt{2}}{3}(1+2\delta)\ ,\\
        N=4:\quad& \beta=2\sqrt{2}\delta\ .\\
    \end{aligned}
    \quad\iff\quad 
    \begin{aligned}
        N=2:\quad& p_{\text{win}}^\text{CHSH} \leq \frac{1}{4}\left(2+\sqrt{1+\delta^2}\right)\ ,\\
        N=3:\quad& p_{\text{win}}^\text{CHSH} \leq \frac{1}{4}\left(2+\frac{\sqrt{2}}{3}(1+2\delta)\right)\ ,\\
        N=4:\quad& p_{\text{win}}^\text{CHSH} \leq \frac{1}{4}\left(2+\sqrt{2}\delta\right)\ .\\
    \end{aligned}
\end{equation}
For a given set of state preparations with a pair-wise overlap bounded by $\delta$, one can derive the maximum success probability of local state discrimination $p_s^{\text{L}}$, which is bounded by
\begin{align}
p_{s}^{N}(\delta)\!=\!\frac{1}{N^2
}\left(\sqrt{1+(N-1)\delta}+(N-1)\sqrt{1-\delta}\right)^2 ,
\end{align}
as pointed out in the main text. In \figref{fig:tradeoff_many} we show the bounds on the CHSH winning probability given the maximum success probability of local state discrimination. Whenever the number of state preparations exceeds two with qubit local dimension we observe a gap between CHSH violation and full state distinguishability regime. Thus, if $N>2$, ensemble non-locality is not feasible as soon as state preparations are not equivalent. 

\begin{figure*}
    \centering
    \includegraphics[width=\textwidth]{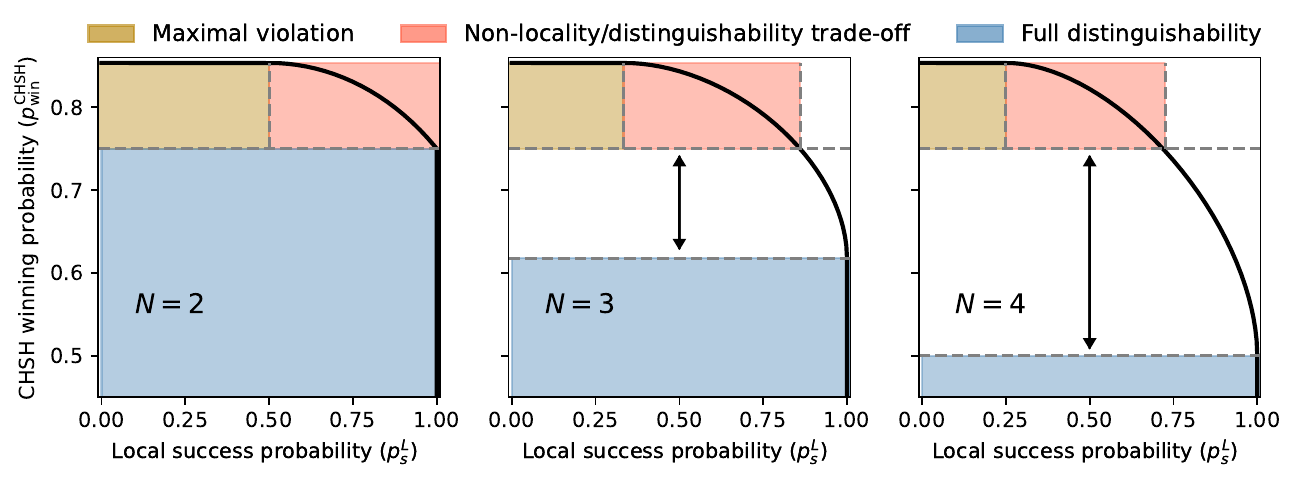}
    \caption{Trade-off between CHSH violation and local state distinguishability characterized through the state discrimination success probability. Here we show cases with $N=2,3,4$ state preparations with local qubit dimension.}
    \label{fig:tradeoff_many}
\end{figure*}

\section{Properties of axisymmetric states}\label{ap:axisym}
In this part of the supplemental material we prove the  properties  used in the main text associated with the family of states in \eqref{eq:qutrit_states}. Concretely,
\begin{align}
    \ket{\psi_z}\! =\! \sqrt{\frac{1\!+\!(N\!-\!1)\delta}{N}}\ket{\xi_0} + \sqrt{\frac{1-\delta}{N}}\sum_{j=1}^{N-1}e^{i\frac{2\pi jz}{N}} \ket{\xi_j}\ ,
\end{align}
with 
\begin{align}
    \ket{\xi_{n}} = \frac{1}{\sqrt{N}}\sum_{j}e^{i\frac{2\pi jn}{N}}\ket{j}\otimes\ket{j} \ .
\end{align}
We first notice that for any $z\neq z'$, the product $\braket{\psi_z|\psi_{z'}}=\delta$. Indeed, since $\braket{\xi_j|\xi_k}=\delta_{jk}$,
\begin{equation}
    \braket{\psi_z|\psi_{z'}}=\frac{1\!+\!(N\!-\!1)\delta}{N}+\frac{1-\delta}{N}\sum_{j=1}^{N-1}e^{i\frac{2\pi j}{N}\left(z-z'\right)}=\delta+\left(1-\delta\right)\left[\frac{1}{N}\sum_{j=0}^{N-1}e^{i\frac{2\pi j}{N}\left(z-z'\right)}\right]\ .
\end{equation}
We identify the last term with the discrete Fourier transform (DFT), hence yielding
\begin{equation}
    \braket{\psi_z|\psi_{z'}}=\delta+\left(1-\delta\right)\delta_{z,z'}\ .
\end{equation}
%In particular, for $z=z'$, $\braket{\psi_z}=1$ while for $z\neq z'$, $\braket{\psi_z}{\psi_{z'}}=\delta$. 
Let us now see the expression of the density matrix $\rho_z=\ketbra{\psi_z}$, from which we will derive the ensemble $\rho=\frac{1}{N}\sum_{z=0}^{N-1}\rho_z$. Namely,
\begin{equation}
\begin{aligned}
    \rho_z=&\frac{1\!+\!(N\!-\!1)\delta}{N}\ketbra{\xi_0}+\sum_{j=1}^{N-1}\sqrt{\frac{1\!+\!(N\!-\!1)\delta}{N}}\sqrt{\frac{1-\delta}{N}}e^{-i\frac{2\pi j z}{N}}\ketbra{\xi_0}{\xi_j}\\
    &+\sum_{j=1}^{N-1}\sqrt{\frac{1\!+\!(N\!-\!1)\delta}{N}}\sqrt{\frac{1-\delta}{N}}e^{i\frac{2\pi j z}{N}}\ketbra{\xi_j}{\xi_0} +\sum_{j,k=1}^{N-1}\frac{1-\delta}{N}e^{i\frac{2\pi z}{N}\left(j-k\right)}\ketbra{\xi_j}{\xi_k}\ .
\end{aligned}
\end{equation}
Performing now the sum over $z$,
\begin{equation}
\begin{aligned}
    \sum_{z=0}^{N-1}\rho_z=&\left(1\!+\!(N\!-\!1)\delta\right)\ketbra{\xi_0}+\sum_{j=1}^{N-1}\sqrt{1\!+\!(N\!-\!1)\delta}\sqrt{1-\delta}\left[\frac{1}{N}\sum_{z=0}^{N-1} e^{i\frac{2\pi z}{N}\left(0-j\right)}\right]\ketbra{\xi_0}{\xi_j}\\
    &+\sum_{j=1}^{N-1}\sqrt{1\!+\!(N\!-\!1)\delta}\sqrt{1-\delta}\left[\frac{1}{N}\sum_{z=0}^{N-1} e^{i\frac{2\pi z}{N}\left(j-0\right)}\right]\ketbra{\xi_j}{\xi_0} +\sum_{j,k=1}^{N-1}\left(1-\delta\right)\left[\frac{1}{N}\sum_{z=0}^{N-1}e^{i\frac{2\pi z}{N}\left(j-k\right)}\right]\ketbra{\xi_j}{\xi_k}\ ,
\end{aligned}
\end{equation}
and identifying the corresponding DFTs,
\begin{equation}
    \sum_{z=0}^{N-1}\rho_z=\left(1\!+\!(N\!-\!1)\delta\right)\ketbra{\xi_0}+\sum_{j=1}^{N-1}\left(1-\delta\right)\ketbra{\xi_j}{\xi_j}\ .
\end{equation}
Thus, the density matrix is diagonal in the Bell basis,
\begin{equation}
   \rho= \frac{1}{N}\sum_{z=0}^{N-1}\rho_z=\frac{1\!+\!(N\!-\!1)\delta}{N}\ketbra{\xi_0}+\sum_{j=1}^{N-1}\frac{1-\delta}{N}\ketbra{\xi_j}{\xi_j}\ ,
\end{equation}
with non-zero eigenvalues $\frac{1\!+\!(N\!-\!1)\delta}{N}$ and $\frac{1-\delta}{N}$. Notice that any other set of normalised vectors $\{\ket{\psi_z}\}$ satisfying $\braket{\psi_z|\psi_{z'}}=\delta$ for $z\neq z'$ are obtained by a unitary transformation $U$ of the axisymmetric family and therefore the corresponding density matrix $U\rho U^\dagger$ will share the same spectrum. Namely, any global restriction must correspond to an \textit{absolute} property of the degenerated spectrum $\left\{\frac{1\!+\!(N\!-\!1)\delta}{N}, \frac{1-\delta}{N},0\right\}$.

We now turn towards the explicit form of local state $\rho^{A}_{z}=\Tr_B \left[\rho_z\right]$ with respect to Alice (and same with Bob),
\begin{equation}
    \begin{aligned}
    \rho^{A}_{z}=&\frac{1\!+\!(N\!-\!1)\delta}{N}\Tr_B \left[\ketbra{\xi_0}\right]+\sum_{j=1}^{N-1}\sqrt{\frac{1\!+\!(N\!-\!1)\delta}{N}}\sqrt{\frac{1-\delta}{N}}e^{-i\frac{2\pi j z}{N}}\Tr_B \left[\ketbra{\xi_0}{\xi_j}\right]\\
    &+\sum_{j=1}^{N-1}\sqrt{\frac{1\!+\!(N\!-\!1)\delta}{N}}\sqrt{\frac{1-\delta}{N}}e^{i\frac{2\pi j z}{N}}\Tr_B \left[\ketbra{\xi_j}{\xi_0}\right] +\sum_{j,k=1}^{N-1}\frac{1-\delta}{N}e^{i\frac{2\pi z}{N}\left(j-k\right)}\Tr_B \left[\ketbra{\xi_j}{\xi_k}\right]\ .
    \end{aligned}
\end{equation}
Note that the element $\Tr_B \left[\ketbra{\xi_j}{\xi_k}\right]$, for $j,k=0,\dots,N-1$ is,
\begin{equation}
    \begin{aligned}
    \Tr_B \left[\ketbra{\xi_j}{\xi_k}\right]=\frac{1}{N}\sum_{m,n=0}^{N-1}e^{i\frac{2\pi}{N}\left(jm-kn\right)}\Tr_B \left[\ketbra{m m}{n n}\right]=\frac{1}{N}\sum_{m=0}^{N-1}e^{i\frac{2\pi}{N}\left(jm-km\right)}\ketbra{m}.
    \end{aligned}
\end{equation}
which translates to
\begin{equation}
    \begin{aligned}
    \rho^{A}_{z}=&\frac{1\!+\!(N\!-\!1)\delta}{N}\frac{1}{N}\sum_{m=0}^{N-1}\ketbra{m}\\
    &+\sqrt{\frac{1\!+\!(N\!-\!1)\delta}{N}}\sqrt{\frac{1-\delta}{N}}\sum_{m=0}^{N-1}\left[\frac{1}{N}\sum_{j=1}^{N-1}e^{-i\frac{2\pi j}{N}\left(z+m\right)}\right]\ketbra{m}\\
    &+\sqrt{\frac{1\!+\!(N\!-\!1)\delta}{N}}\sqrt{\frac{1-\delta}{N}}\sum_{m=0}^{N-1}\left[\frac{1}{N}\sum_{j=1}^{N-1}e^{i\frac{2\pi j}{N}\left(z+m\right)}\right]\ketbra{m}\\
    &+\left(1-\delta\right)\sum_{m=0}^{N-1}\left[\frac{1}{N}\sum_{j=1}^{N-1}e^{i\frac{2\pi j}{N}\left(z+m\right)}\right]\left[\frac{1}{N}\sum_{j=1}^{N-1}e^{-i\frac{2\pi k}{N}\left(z+m\right)}\right]\ketbra{m}\ .
    \end{aligned}
\end{equation}
Replacing the DFTs one recovers
\begin{equation}
    \begin{aligned}
    \rho^{A}_{z}=&\frac{1\!+\!(N\!-\!1)\delta}{N}\frac{1}{N}\sum_{m=0}^{N-1}\ketbra{m}\\
    &+\sqrt{\frac{1\!+\!(N\!-\!1)\delta}{N}}\sqrt{\frac{1-\delta}{N}}\sum_{m=0}^{N-1}\left[\delta_{m,N-z}-\frac{1}{N}\right]\ketbra{m}\\
    &+\sqrt{\frac{1\!+\!(N\!-\!1)\delta}{N}}\sqrt{\frac{1-\delta}{N}}\sum_{m=0}^{N-1}\left[\delta_{m,N-z}-\frac{1}{N}\right]\ketbra{m}\\
    &+\left(1-\delta\right)\sum_{m=0}^{N-1}\left[\delta_{m,-z}-\frac{1}{N}\right]\left[\delta_{m,N-z}-\frac{1}{N}\right]\ketbra{m}\ ,
    \end{aligned}
\end{equation}
where with our notation $N-0\equiv 0$. Simplifying,
\begin{equation}
    \begin{aligned}
    \rho^{A}_{z}=&\left[\frac{1\!+\!(N\!-\!1)\delta}{N}-2\sqrt{\frac{1\!+\!(N\!-\!1)\delta}{N}}\sqrt{\frac{1-\delta}{N}}+\frac{1-\delta}{N}\right]\left[\frac{1}{N}\sum_{m=0}^{N-1}\ketbra{m}\right]\\
    &+\left[\left(1-\delta\right)+2\sqrt{\frac{1\!+\!(N\!-\!1)\delta}{N}}\sqrt{\frac{1-\delta}{N}}-2\frac{1-\delta}{N}\right]\ketbra{N-z}\\ 
    =&\frac{1}{N^2
    }\left(\sqrt{1\!+\!(N\!-\!1)\delta}-\sqrt{1-\delta}\right)^2\sum_{m\neq N-z}\ketbra{m}\\
    &+\frac{1}{N^2
    }\left(\sqrt{1+(N-1)\delta}+(N-1)\sqrt{1-\delta}\right)^2\ketbra{N-z}\ .
    \end{aligned}
\end{equation}
One can simply write 
\begin{equation}
    \rho^{A}_{z}=p_s^N\ketbra{N-z}+p_e^N\sum_{m\neq N -z}\ketbra{m}
\end{equation}
with 
\begin{equation}
    p_s^N=\frac{1}{N^2
    }\left(\sqrt{1+(N-1)\delta}+(N-1)\sqrt{1-\delta}\right)^2, \quad p_e^N=\frac{1}{N^2
    }\left(\sqrt{1\!+\!(N\!-\!1)\delta}-\sqrt{1-\delta}\right)^2\ ,
\end{equation}
thus recovering the properties listed in the main text.

\section{Fidelity bound}\label{ap:fidbound}
In this part of the supplemental material we show that given a set of $N$ normalised states $\{\ket{\psi_z}\}$ such that $\abs{\braket{\psi_z|\psi_{z'}}}\leq\delta$ for $z\neq z'$, the maximal fidelity of $\rho=\frac{1}{N}\sum_{z=0}^{N-1}\ketbra{\psi_z}$ with any other state $\ket{\varphi}$ verifies 
\begin{equation}
    f_{\varphi}(\rho)=\bra{\varphi}\rho\ket{\varphi}\leq \frac{1+(N-1)\delta}{N}\ ,
\end{equation}
with the equality holding if and only if $\braket{\psi_z|\psi_{z'}}=\delta$ and $\ket{\varphi}\propto\sum_z \ket{\psi_z}$.
To see this, firstly note that
\begin{equation}
    f_{\varphi}(\rho)=\bra{\varphi}\rho\ket{\varphi}\leq\lambda_{\max}(\rho)\ .
\end{equation}
The largest eigenvalue of $\rho$ is bounded from above by \cite{WOLKOWICZ1980471},
\begin{equation}
    \lambda_{\max}(\rho)\leq \frac{\Tr\left[\rho\right]}{N}+\sqrt{\frac{N-1}{N}}\sqrt{\Tr\left[\rho^2\right]-\frac{\Tr\left[\rho\right]^2}{N}}\ .
\end{equation}
Since $\Tr\left[\rho\right]=1$ and $\Tr\left[\rho^2\right]\leq \frac{1}{N}+\frac{N-1}{N}\delta^2$ due to $\abs{\braket{\psi_z|\psi_{z'}}}\leq\delta$, the upper bound reads
\begin{equation}
    \lambda_{\max}(\rho)\leq \frac{1}{N}+\sqrt{\frac{N-1}{N}}\sqrt{\Tr\left[\rho^2\right]-\frac{1}{N}}\leq\frac{1}{N}+\sqrt{\frac{N-1}{N}}\sqrt{\frac{N-1}{N}\delta^2}=\frac{1+(N-1)\delta}{N}\ .
\end{equation}
Hence, 
\begin{equation}
    f_{\varphi}(\rho)\leq \frac{1+(N-1)\delta}{N}\ .
\end{equation}
The equality holds whenever the bounds over $\Tr\left[\rho^2\right]$ and $\lambda_{\max}(\rho)$ are tight, and $\ket{\varphi}$ coincides with the eigenvector associated with the largest eigenvalue of $\rho$. The first one takes place if $\braket{\psi_z|\psi_{z'}}=\delta$, which in turn implies that the $N-1$ non-zero smallest eigenvalues of $\rho$ are all equal. Actually, this last condition is the only requirement for the bound on $\lambda_{\max}(\rho)$ to be saturated \cite{WOLKOWICZ1980471}. Finally, the eigenvector associated with $\lambda_{\max}(\rho)$ is the normalised state $\ket{\phi}=\frac{1}{\sqrt{N+N(N-1)\delta}}\sum_z \ket{\psi_z}$. To see this we just check that $\rho\ket{\phi}=\frac{1+(N-1)\delta}{N}\ket{\phi}$:
\begin{equation}
\begin{aligned}
    \rho\ket{\phi}=&\ \frac{1}{N}\frac{1}{\sqrt{N+N(N-1)\delta}}\sum_{z,z'}\braket{\psi_z|\psi_{z'}}\ket{\psi_z}=\frac{1}{N}\frac{1}{\sqrt{N+N(N-1)\delta}}\left[\sum_{z}\ket{\psi_z}+(N-1)\sum_{z}\delta \ket{\psi_z}\right]\\
    =&\ \frac{1+(N-1)\delta}{N}\frac{1}{\sqrt{N+N(N-1)\delta}}\sum_{z}\ket{\psi_z}= \frac{1+(N-1)\delta}{N}\ket{\phi}\ .
\end{aligned}
\end{equation}

%\bibliography{main}% Produces the bibliography via BibTeX. 

\end{widetext}

\end{document}